\documentclass[numsec,webpdf,modern,medium,namedate]{oup-authoring-template}

\onecolumn

\graphicspath{{Fig/}}

\theoremstyle{thmstyleone}%
\theoremstyle{thmstyletwo}%
\theoremstyle{thmstylethree}%

\usepackage{setspace}
\usepackage[fontsize=11pt]{fontsize}

\usepackage{titlesec}

\titlespacing*{\section}
  {0pt}                              
  {1ex plus 0.3ex minus 0.2ex}     
  {0.7ex}                            

\titlespacing*{\subsection}
  {0pt}
  {1.2ex plus 0.25ex minus 0.15ex}
  {0.5ex}

\titlespacing*{\subsubsection}
  {0pt}
  {1.0ex plus 0.2ex minus 0.1ex}
  {0.4ex}

\titleformat{\section}
  {\normalfont\large\bfseries}
  {\thesection.}{0.6em}{}

\titleformat{\subsection}
  {\normalfont\normalsize\bfseries}
  {\thesubsection}{0.6em}{}

\titleformat{\subsubsection}
  {\normalfont\normalsize\bfseries}
  {\thesubsubsection}{0.6em}{}

\newcommand{\bx}{\mathbf{x}}
\newcommand{\bs}{\mathbf{s}}
\newcommand{\bbeta}{\boldsymbol{\beta}}
\newcommand{\bW}{\boldsymbol{W}}
\newcommand{\bX}{\mathbf{X}}
\newcommand{\bz}{\boldsymbol{z}}
\newcommand{\bg}{\boldsymbol{g}}

\begin{document}

\journaltitle{Journals of the Royal Statistical Society}
\DOI{DOI HERE}
\copyrightyear{XXXX}
\pubyear{XXXX}
\access{Advance Access Publication Date: Day Month Year}
\appnotes{Original article}

\firstpage{1}



\title[Modelling U.S.\ School Gun Violence]{Kernel Regression for Spatial and Spatio-temporal Residual Risk: Application to School Shootings in the Contiguous United States}

\author[1,$\ast$]{Tilman M.\ Davies\ORCID{0000-0003-0565-1825}}
\author[2]{Michael R.\ Desjardins\ORCID{0000-0002-2789-5460}}
\author[3]{Alexander Hohl\ORCID{0000-0002-9103-0744}}
\author[4]{Guangzhen Wu\ORCID{0000-0001-5767-0986}}

\authormark{Davies et al.}

\address[1]{\orgdiv{Department of Mathematics \& Statistics}, \orgname{University of Otago}, \orgaddress{\street{Dunedin}, \country{New Zealand}}}
\address[2]{\orgdiv{Department of Epidemiology}, \orgname{Johns Hopkins Bloomberg School of Public Health}, \orgaddress{\street{Baltimore}, \state{Maryland}, \country{USA}}}
\address[3]{\orgdiv{Department of Geography}, \orgname{University of Utah}, \orgaddress{\street{Salt Lake City}, \state{Utah}, \country{USA}}}
\address[4]{\orgdiv{Department of Criminology \& Sociology}, \orgname{University of Utah}, \orgaddress{\street{Salt Lake City}, \state{Utah}, \country{USA}}}

\corresp[$\ast$]{Address for correspondence. Assoc.\ Prof.\ T.\ M.\ Davies, University of Otago, Dunedin, 9016, New Zealand. \href{Email:email-id.com}{tilman.davies@otago.ac.nz}}

\received{Date}{0}{Year}
\revised{Date}{0}{Year}
\accepted{Date}{0}{Year}

\abstract{School gun violence in the United States is a complex phenomenon spanning social, epidemiological, demographic, and political dimensions. It remains unclear where incidents are unusually concentrated nationally after accounting for the distribution and characteristics of schools. Using a newly linked case-control dataset comprising 959 gun-violence incidents at public K-12 schools in the contiguous United States during 2000-2024, we develop a semiparametric kernel-regression framework combining school-level predictors with spatial and continuously evolving spatio-temporal residual structure. Fisher-weighted orthogonalisation defines how predictor-aligned variation is allocated between fixed and smooth components, while repeated control sampling and Monte Carlo reassignment support stable mapping and local exceedance assessment. The models identify stable school-level associations, including substantially higher adjusted odds for larger, middle, and high schools, while revealing residual structure beyond the background distribution of schools. Elevated residual odds become concentrated in a broad central-eastern corridor from the mid-2010s onward, with the strongest evidence in recent years. The analysis offers both statistical and application-specific insights. Statistically, it shows how covariate-adjusted residual surfaces can characterise local departures in case-control processes evolving over space and time. For the application, it provides epidemiological clues identifying regions in which broader social, policy, and environmental conditions may warrant targeted investigation.}
\keywords{case-control, gun violence, point pattern, semiparametric model, smoothing}


\singlespacing
\maketitle
\singlespacing

\section{Introduction}\label{sec:intro}

Although by no means exclusive to the United States, gun-related violence at schools is an especially prominent concern in a country with exceptionally high rates of firearm violence relative to other high-income nations \citep{richardson2011homicide}. Its national geography remains difficult to characterise. Incidents are rare, heterogeneous, and not recorded through a single unified official reporting system; instead, available compilations differ in their definitions, scope, and collection strategies \citep{nowicki2020gao,stewart2022school,comer2024definitional}. Public understanding is also disproportionately shaped by a small number of highly publicised mass-casualty events, although the broader phenomenon includes incidents of widely varying severity, motivation, and circumstance \citep{kolbe2020school,riedman2025ssdb}. Consequently, basic national questions remain incompletely answered: where is school gun violence unusually concentrated, to what extent do apparent concentrations simply reflect the distribution and characteristics of schools, and how has this geography changed over time?

Existing national studies have illuminated several important aspects of school gun violence, but do not directly answer these questions. Recent work has examined incident severity, event characteristics, school-district context, and state firearm-law environments \citep{fridel2021contextual,gammell2022descriptive,joseph2023role,wippell2026preventable}. These studies provide valuable evidence on individual incidents, contextual correlates, and broad policy associations, but have largely relied on descriptive summaries, incident-level regression, or analyses aggregated to districts, counties, or states. There remains, consequently, little evidence on how the adjusted risk of school gun violence varies continuously across the country, or on how that geography has evolved through time.

The remaining challenge is statistical as well as substantive. Apparent concentrations of incidents may simply reflect the geographical distribution of schools and students, while measured school characteristics may explain further variation in incident odds. To separate any such concentration from the underlying distribution of schools and students, kernel relative-risk estimation provides a flexible means of comparing the spatial distributions of cases and controls, producing a continuously varying measure of case concentration relative to an appropriate background population \citep{bithell1991estimation,kelsall1995kernel,davies2018tutorial}. Related constructions extend naturally to space and time \citep{fernando2014generalizing}. Classical relative-risk estimators, however, do not readily accommodate record-level predictors. \citet{keldig:1998} addressed this limitation by embedding a kernel-smoothed spatial component within a semiparametric binary regression model, allowing measured covariates and localised residual structure to be represented jointly.

We use this framework to analyse a newly linked national case-control dataset of school gun-violence incidents and contemporaneous school records. Building on the semiparametric kernel-regression model of \citet{keldig:1998}, we refine the separation between measured school characteristics and remaining local structure, and extend the approach from purely spatial analysis to residual risk evolving continuously over space and time. A focused simulation study examines the consequences of this modelling choice, while repeated control sampling and Monte Carlo assessment are used to evaluate the stability and strength of the patterns identified in the application.

The balance of the paper is structured as follows. Section~\ref{sec:data} describes the incident data, school-record linkage, control-sampling design, and predictors. Section~\ref{sec:model} presents the spatial and spatio-temporal models and their implementation. Section~\ref{sec:sim} reports the simulation study, Section~\ref{sec:res} presents the application results, and Section~\ref{sec:disc} concludes with discussion and limitations.

\section{School Gun Violence Data and Study Design}\label{sec:data}

\subsection{Incident selection and school-record linkage}\label{sec:casedata}

Our incident data are drawn from the K-12 School Shooting Database (SSDB); see \cite{riedman2025ssdb}, from which we extract qualifying firearm-related incidents at K-12 schools across the United States. Details of the SSDB's data collection and verification procedures are available at \footnotesize\url{https://k12ssdb.org}\normalsize. The database contains a wide range of variables describing each recorded incident, which we use to define the analytical case set.

We scrutinise the 25-year period from 1 January 2000 to 31 December 2024, comprising 2,247 recorded events. Although we adopt a relatively broad definition of a `school gun violence event', our interest lies in incidents involving school students and/or staff on school property in which there was intent to harm one or more individuals other than the perpetrator, whether in a premeditated or spontaneous fashion. We  therefore exclude events classified as accidental, incidents involving an isolated suicide attempt, and records with a low reliability indicator (qualifying incidents must have been the subject of at least one news article with a named author). This initial filtering exercise leaves 1,176 records.

The SSDB provides the incident date and location, school name, city and state, and school level. We simplify the latter to four categories: Primary/Elementary, Middle/Junior High, High/Senior High, and Other, with the final category reserved for schools that cannot readily be assigned to one of the preceding groups. Incident locations are supplied as WGS84 longitude and latitude coordinates (EPSG:4326). These are projected using the NAD83/Conus Albers Equal Area system (EPSG:5070) and rescaled from metres to kilometres for analysis.

Although the SSDB contains detailed information on the incidents themselves, it includes comparatively little information on the corresponding schools. We therefore obtain contemporaneous school descriptors from the National Center for Education Statistics (NCES), part of the United States Department of Education. Data are drawn from the NCES Elementary/Secondary Information System (ElSi), wherein schools possess unique 12-digit identifiers and are represented by school-year-specific records containing enrolment and other descriptive information.

Because these identifiers are not recorded in the SSDB, candidate matches are identified using school name and broad location, including state and city, and are then checked against school level, ZIP code, and geographical coordinates. The matching exercise is complicated by the absence of standardised school naming conventions and by occasional data-entry errors in both sources. We exclude records with unresolved links, incompatible school classifications, invalid or internally inconsistent enrolment values, or erroneous geographical coordinates. We also omit private schools because they account for relatively few qualifying incidents and their NCES records are less reliable over the study period than those for public schools; indeed, NCES/ElSi ceased hosting private-school descriptors after 2021.

\begin{figure}[h!]
    \centering
    \includegraphics[width=0.48\linewidth]{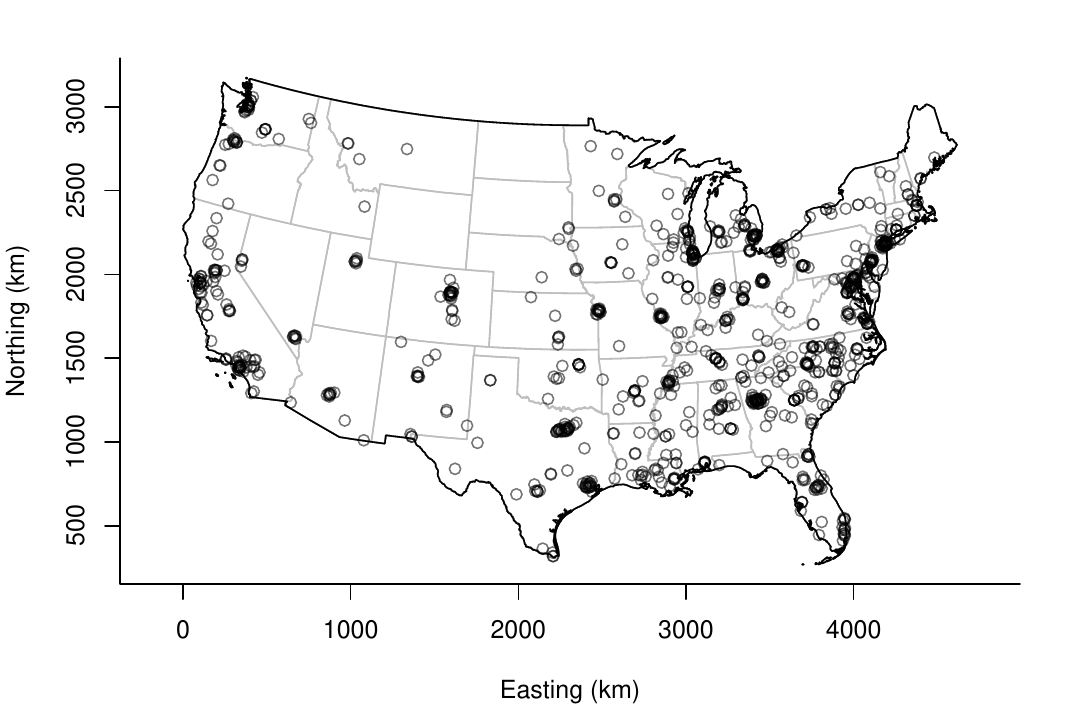}
    \includegraphics[width=0.48\linewidth]{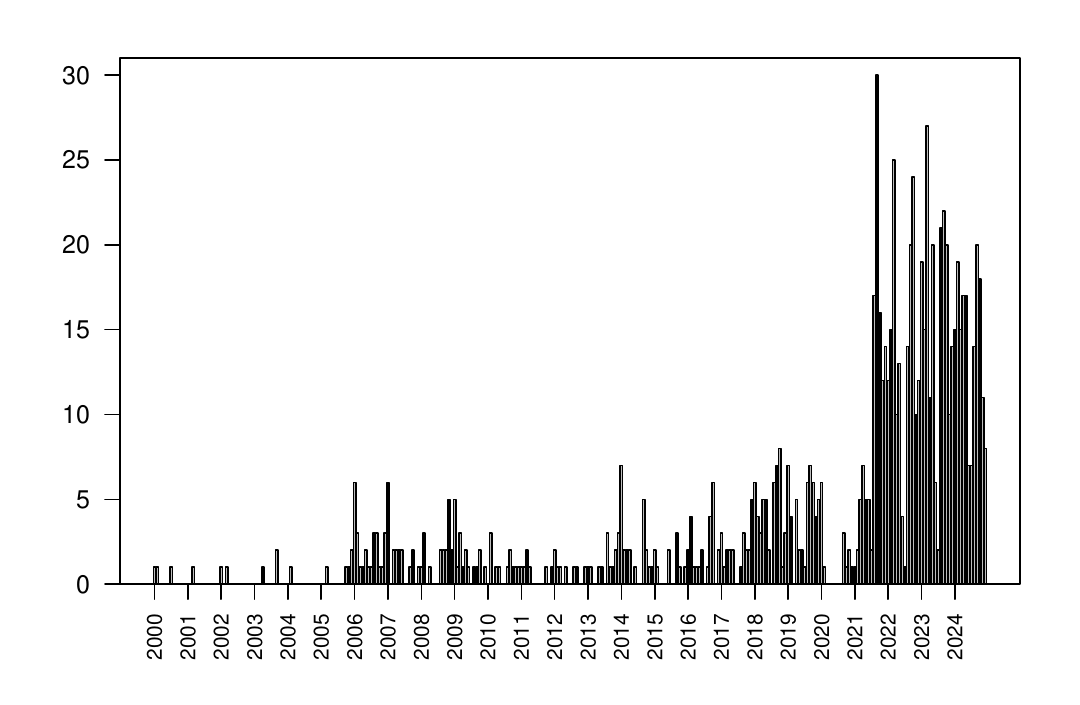}
    \caption{Locations (left) and monthly counts (right) of the 959 qualifying gun-violence incidents at public K-12 schools in the contiguous United States, 2000-2024.}
    \label{fig:rawdata}
\end{figure}

The culmination of these efforts resulted in an analytical case set comprising 959 NCES-matched gun-violence incidents at public K-12 schools in the contiguous United States. Their spatial and temporal margins are shown in Figure~\ref{fig:rawdata}, which reveals substantial geographical unevenness and a marked increase in recorded incidents during the latter years of the study period. A lull is visible during 2020, coinciding with widespread pandemic-related disruption to in-person schooling, followed by a pronounced increase from 2021 onwards. Neither the spatial concentrations nor the temporal increase can, however, be interpreted directly as elevated risk without comparison to the changing distribution and characteristics of schools.

\subsection{Control selection and repeated sampling}\label{sec:condata}

Control records are drawn from the full set of NCES public-school records for the 48 contiguous states over the corresponding period from 2000 to 2024. Following removal of fully virtual schools and records with missing, incorrect, or internally inconsistent enrolment information, the eligible control pool contains over two million school-year records. This is a combined stack of annual data, so most individual schools appear in multiple years, providing contemporaneous snapshots of their enrolment and demographic characteristics. 

For each analysis, we draw $c=10$ controls per case, representing a pragmatic trade-off between estimate stability and computational burden. Controls are frequency-matched to the cases by calendar year. Specifically, if year $t$ contains $n^{(t)}$ incidents, we sample without replacement $cn^{(t)}$ eligible control school-year records from that same year, after excluding NCES identifiers associated with the cases. The resulting comparison is therefore anchored, within each year, to the population of public schools as it was recorded at that time, rather than borrowing controls from years in which the number and composition of schools may have differed.

This design intentionally absorbs temporal changes in the overall proportion of sampled records classified as cases. The inferential focus is consequently on associations with school characteristics and on relative spatial or spatio-temporal pattern within the year-matched case-control population, rather than on estimating national annual incident prevalence.

Exact incident dates are retained for the cases. For the continuous spatio-temporal analysis, each control school-year record is assigned a reproducible pseudo-date drawn uniformly from within its calendar year. This avoids concentrating all controls at a single arbitrary date while preserving the year-matched sampling design.

Rather than relying on one randomly generated control set, all reported analyses are aggregated over $K=100$ independently sampled sets. This repeated-sampling, or ``bagging'', strategy reduces sensitivity to the particular controls selected in any one draw and allows variability arising from control sampling to contribute to the reported coefficient, surface, and exceedance summaries. A three-dimensional visualisation of the incidents and one representative year-matched control sample is provided in Figure \ref{fig:casecon3d}. An interactive version of this graphic is available both in the supplementary materials and at \footnotesize\url{https://www.stats.otago.ac.nz/research/davies-gv/data.html}\normalsize.

\begin{figure}[h!]
    \centering
    \includegraphics[width=0.65\linewidth]{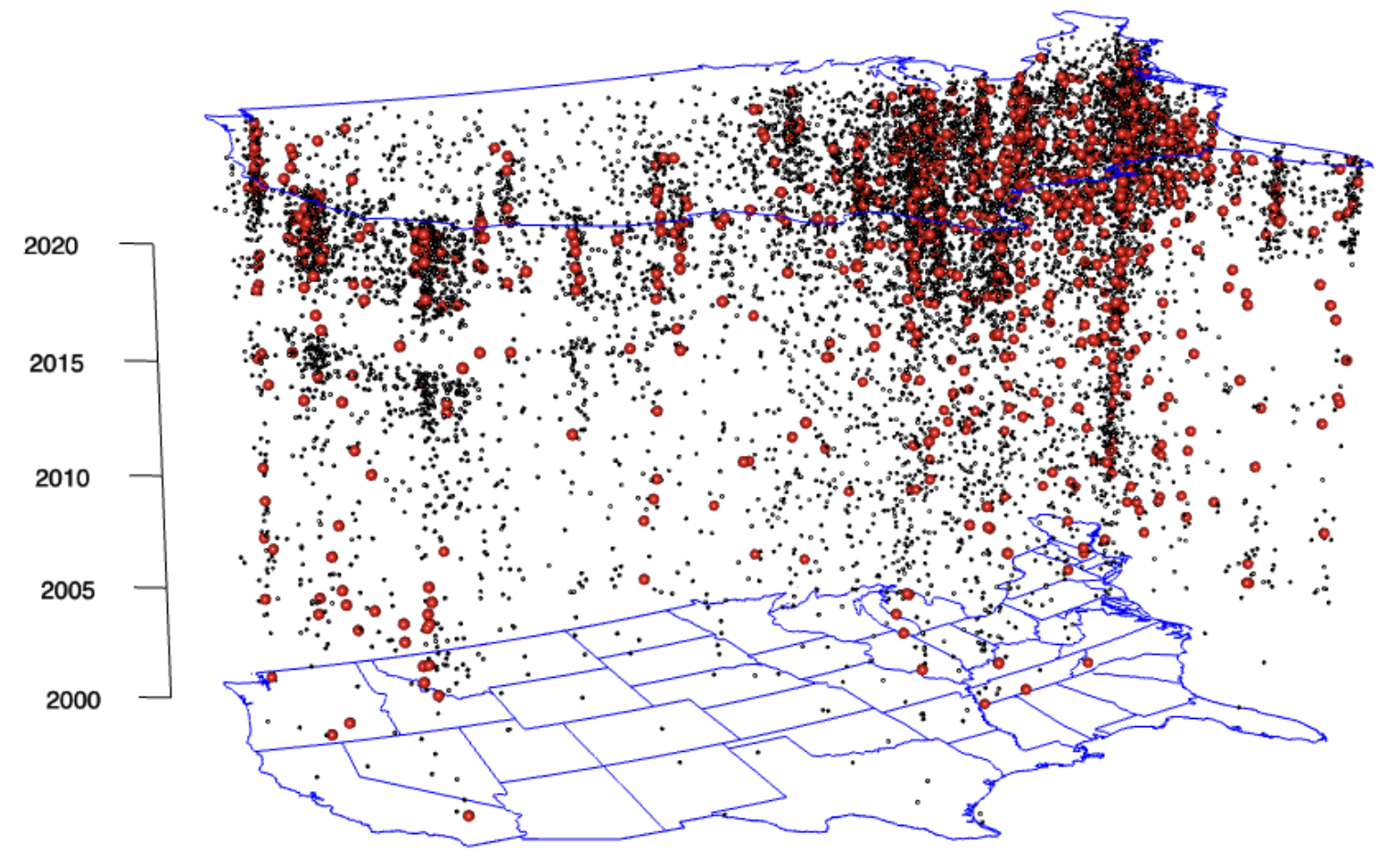}
    \vspace{-2mm}
    \caption{Spatio-temporal plot of the gun-violence incidents (large/red points) alongside one random sample from the NCES/ElSi control pool (small/black points); the controls are frequency-matched by incident year at a ratio of 10:1.}
    \label{fig:casecon3d}
\end{figure}

\subsection{School-level predictors}\label{sec:predictors}

The occurrence of a gun-violence incident may be associated in part with characteristics of the school itself. For example, a school enrolling many hundreds of students may face greater opportunity for an incident than a much smaller otherwise comparable school, purely by virtue of size \citep{fridel2021contextual}. We therefore retrieve a common set of predictors from the contemporaneous NCES record for every case and control.

These comprise total student enrolment, the proportion of enrolled students who are male, the proportion who are White, and school level, classified as Primary, Middle, High, or Other. Total enrolment is log-transformed, with Primary schools used as the reference category for school level. Exploratory analyses indicate curvature in the associations with racial composition and deprivation, so the proportion-White and ADI variables are each ultimately represented quadratically. All numeric predictors are standardised using means and standard deviations calculated from the cleaned eligible NCES school-year population.

We additionally include the Area Deprivation Index (ADI) as a contextual measure of local socioeconomic disadvantage not captured by the NCES school descriptors. Census-tract-level ADI values are derived from United States Census and American Community Survey (ACS) inputs, with larger values indicating greater deprivation, and are assigned to schools according to their geographical location. Because tract-level ADI is not available annually across the full study period, each location is assigned its average value over the available years and this value is carried across school-year records. The corresponding coefficient therefore represents an association with a fixed, long-run tract-level deprivation measure rather than a within-tract temporal effect.

Although ADI is technically piecewise constant over census tracts, this spatial resolution is relatively fine compared with the national scale of the analysis. We do not include county- or state-level predictors in the present school-year model because their substantially coarser support would raise additional issues of spatial alignment and contextual interpretation. Hence, by design, broad remaining spatial and spatio-temporal structure is represented through the smooth component described in the following section.

\section{Models and Implementation}\label{sec:model}

We consider a semiparametric binary regression model in which measured school characteristics enter through a fixed-effects component and remaining local structure is represented through kernel smoothing. Within this additive construction, a central distinction is made between two related but different nonparametric quantities. The first is an observation-level smooth contribution used jointly with the fixed effects to fit the binary model, and is constrained to be orthogonal to the columns of the observed design matrix under the current Fisher weights. The second is a continuous spatial or spatio-temporal surface used to map residual risk after fitting. Because the school-level covariates are not defined at every map location, this continuous surface cannot in general satisfy the same full orthogonality constraint. We define both quantities explicitly below and clarify their respective interpretations.

A natural alternative to the kernel representation would be to model the smooth spatial contribution using the penalised spline or basis-expansion methods commonly employed in generalised additive modelling. We adopt a kernel construction because the scientific target is naturally expressed as a local case-control comparison: it extends classical kernel relative-risk estimation directly to adjustment for record-level predictors, gives the bandwidth an interpretable geographical meaning, and provides a common local formulation for model fitting, continuous mapping, and extension through time. This choice should not be interpreted as a claim that spline-based smoothers are intrinsically unsuitable for these data. Rather, it preserves a direct connection between the fitted smooth and the relative-risk construction motivating the analysis, while the orthogonalisation described below defines explicitly how variation aligned with the measured predictors is allocated between the model components.

In what follows, suppose we possess $n=n_0+n_1$ records, where $y_i\in\{0,1\}$ identifies record $i$ as a control or case, respectively. Let $\bx_i\in\mathbb{R}^{p+1}$ contain an intercept and the $p$ measured predictors, and write $\bX=[\bx_1,\ldots,\bx_n]^\top$ for the corresponding $n\times(p+1)$ design matrix. The planar location of record $i$ is denoted by $\bs_i\in\mathcal W\subset\mathbb{R}^2$.

\subsection{Purely spatial model}\label{sec:spatial}

\subsubsection{Model formulation}

Let $p_i=\Pr(y_i=1\mid\bx_i,\bs_i)$. At the observed records, we consider the semiparametric representation
\begin{equation}\label{eq:M1def}
\operatorname{logit}(p_i)=\bx_i^\top\bbeta+g_{\perp}(\bs_i),\qquad i=1,\ldots,n,
\end{equation}
where $\bbeta=[\beta_0,\ldots,\beta_p]^\top$ contains the fixed-effects coefficients and $g_{\perp}(\bs_i)$ denotes an $\bX$-orthogonal smooth contribution at the $i$th observed location. The notation $g_{\perp}(\bs_i)$ is used here specifically for the values entering the fitted model at the observed records; the separate continuous surface used for mapping is defined in Section~\ref{sec:infer}.

To motivate the smooth component, let $f_1(\bs)$ and $f_0(\bs)$ denote the spatial densities of case and control locations, respectively. A natural measure of relative case concentration is the density ratio
\begin{equation}\label{eq:denrat}
r(\bs)=\frac{f_1(\bs)}{f_0(\bs)},
\end{equation}
or, on the logarithmic scale, $\log r(\bs)=\log f_1(\bs)-\log f_0(\bs)$.
Positive values of $\log r(\bs)$ indicate locations at which cases are relatively more concentrated than controls, while negative values indicate comparatively lower concentration.

The present model retains this local case-control comparison while allowing adjustment for measured school-level predictors. For a fixed $\bbeta$, define the provisional kernel component $\widetilde g_h(\bs)$ at each location $\bs\in\mathcal W$ as the value satisfying the kernel-weighted local score equation
\begin{equation}\label{eq:M1_local_score}
\sum_{i=1}^n\kappa_h(\|\bs-\bs_i\|)\left[y_i-\operatorname{logit}^{-1}\{\bx_i^\top\bbeta+\widetilde g_h(\bs)\}\right]=0,
\end{equation}
where $\kappa_h(\bs)=h^{-2}\kappa(h^{-1}\bs)$ is a radially symmetric spatial kernel, with $\kappa:\mathbb R^2\to\mathbb R$ and scalar bandwidth $h$. Thus, $\widetilde g_h(\bs)$ is defined through a local kernel comparison of cases and controls after adjustment for the linear predictor $\bx_i^\top\bbeta$. It is provisional because its values at the observed records have not yet been subjected to the orthogonality constraint defining $g_{\perp}(\bs_i)$ in \eqref{eq:M1def}.

The connection with the classical density-ratio construction \eqref{eq:denrat} is clearest in the intercept-only case, for which \eqref{eq:M1_local_score} becomes
\[
\sum_{i=1}^n\kappa_h(\|\bs-\bs_i\|)\left[y_i-\operatorname{logit}^{-1}\{\beta_0+\widetilde g_h(\bs)\}\right]=0.
\]
Solving for the local fitted odds gives
\[
\exp\{\beta_0+\widetilde g_h(\bs)\}=\frac{\sum_{i:y_i=1}\kappa_h(\|\bs-\bs_i\|)}{\sum_{i:y_i=0}\kappa_h(\|\bs-\bs_i\|)}.
\]
Hence $\beta_0+\widetilde g_h(\bs)$ is the logarithm of the ratio of kernel-weighted case mass to kernel-weighted control mass. Expressed in terms of kernel density estimators, this gives
\[
\beta_0+\widetilde g_h(\bs)=\log\left\{\frac{\widehat f_1(\bs)}{\widehat f_0(\bs)}\right\}+\log\left(\frac{n_1}{n_0}\right).
\]
Thus, apart from the additive constant absorbed by the intercept, the provisional smooth coincides with the classical kernel log-density-ratio estimator of relative risk. This provides the formal link motivating the kernel construction: in the absence of covariates, it reduces directly to the familiar case-control relative-risk estimator, while in the general model it extends the same local comparison to allow adjustment for measured predictors.

This construction therefore differs from the penalised spline or basis-expansion representation more commonly used for smooth terms in generalised additive modelling \citep{wood2017gam}. As noted above, the distinction is not that one class of smoother is inherently more suitable than the other, but that $\widetilde g_h(\bs)$ is motivated and defined through a kernel-localised case-control comparison.



\vspace{-0mm}
\subsubsection{Orthogonalised IRLS fitting}\label{sec:imp}

A first-order IRLS approximation to the local score equation \eqref{eq:M1_local_score} leads naturally to a Nadaraya-Watson smoother of the working residuals. The resulting provisional smooth is then orthogonalised under the current Fisher weights before updating the fixed-effects coefficients.

Let $\bg_{\perp}=[\bg_\perp(\bs_1),\ldots,\bg_\perp(\bs_n)]^\top$. At a current iterate $(\bbeta,\bg_{\perp})$, define
\begin{equation}\label{eq:etapw}
\eta_i=\bx_i^\top\bbeta+g_{\perp}(\bs_i),\qquad
p_i=\operatorname{logit}^{-1}(\eta_i),\qquad
w_i=p_i(1-p_i),
\end{equation}
and form the usual working response
\begin{equation}\label{eq:z}
z_i=\eta_i+\frac{y_i-p_i}{w_i}.
\end{equation}
With $\bbeta$ held fixed, let
\begin{equation}\label{eq:u}
u_i=z_i-\bx_i^\top\bbeta
\end{equation}
denote the corresponding partial working residual.

Under the quadratic IRLS approximation, updating the provisional smooth at a location $\bs$ amounts to minimizing a locally weighted least-squares criterion, yielding the Fisher-weighted Nadaraya-Watson estimator
\begin{equation}\label{eq:M1_tildeg}
\widetilde g_h(\bs)=\frac{\sum_{i=1}^n\kappa_h(\|\bs-\bs_i\|)w_i u_i}{\sum_{i=1}^n\kappa_h(\|\bs-\bs_i\|)w_i}.
\end{equation}

Let
\[
\widetilde{\bg}=\left[\widetilde g_h(\bs_1),\ldots,\widetilde g_h(\bs_n)\right]^\top,\qquad
\bW=\operatorname{diag}(w_1,\ldots,w_n).
\]
The orthogonalised observation-level contribution is then obtained by removing the component of $\widetilde{\bg}$ lying in the fixed-effects space under the current Fisher inner product,
\begin{equation}\label{eq:M1_orthog}
\bg_{\perp}
=
\widetilde{\bg}
-
\bX(\bX^\top\bW\bX)^{-1}\bX^\top\bW\widetilde{\bg},
\end{equation}
so that
\[
\bX^\top\bW\bg_{\perp}=\mathbf0.
\]

Given $\bg_{\perp}$, the regression coefficients are updated by
\begin{equation}\label{eq:M1_beta_update}
\bbeta\leftarrow(\bX^\top\bW\bX)^{-1}\bX^\top\bW(\bz-\bg_{\perp}),
\end{equation}
where $\bz=(z_1,\ldots,z_n)^\top$. Starting from the covariate-only logistic fit with $\bg_{\perp}=\mathbf0$, equations \eqref{eq:etapw}-\eqref{eq:M1_beta_update} are iterated until the fitted linear predictor and regression coefficients stabilise.

The projection in \eqref{eq:M1_orthog} defines explicitly how variation shared by the provisional kernel smooth and the observed predictors is allocated. Combining \eqref{eq:M1_orthog} with the additive predictor gives
\[
\bX\bbeta+\widetilde{\bg}
=
\bX\left\{\bbeta+(\bX^\top\bW\bX)^{-1}\bX^\top\bW\widetilde{\bg}\right\}
+
\bg_{\perp},
\]
where $\bg_{\perp}$ is orthogonal to $\bX$ under the Fisher inner product induced by $\bW$. Thus, the component of the provisional smooth aligned with the measured predictors is absorbed into the fixed-effects contribution, rather than remaining available for allocation to the smooth component. Because the Fisher weights are updated throughout the IRLS iterations, the identifying condition applies under the fitted Fisher metric at convergence.

This projection is closely related algebraically to restricted spatial regression \citep{hodges2010adding}. Here, however, its purpose is simply to define the decomposition adopted by the semiparametric kernel model and thereby remove ambiguity in how variation shared between the measured predictors and the smooth component is allocated. In particular, the construction should not be interpreted as a general remedy for omitted spatial confounding, nor does it confer a causal interpretation on the regression coefficients. Rather, the fitted coefficients represent adjusted associations under the declared Fisher-orthogonal decomposition. Computationally, the projection also resolves the allocation in a single step for fixed Fisher weights, avoiding the inner regression-smoothing backfitting loop employed by the original algorithm of \citet{keldig:1998}. The implications of this modelling choice, including comparison with unrestricted kernel backfitting, are examined in Section~\ref{sec:sim}.

\subsubsection{Bandwidth selection}\label{sec:bw}

The IRLS fitting algorithm described in Section~\ref{sec:imp} proceeds conditionally on a fixed kernel bandwidth, which determines the spatial scale at which residual structure is represented. In practice, we estimate this tuning parameter from the data using leave-one-out predictive performance and update it during the early stages of the IRLS iterations.

For a candidate bandwidth $h$ and the current $\bbeta$, the provisional kernel smoother \eqref{eq:M1_tildeg} is evaluated at each observation after omitting that observation from the kernel estimate. This yields the leave-one-out linear predictor
\[
\hat\eta_{-i}(h)
=
\bx_i^\top\bbeta
+
\widetilde g_{h,-i}(\bs_i),
\]
and corresponding fitted probability
\[
\hat p_{-i}(h)
=
\operatorname{logit}^{-1}\{\hat\eta_{-i}(h)\},
\]
where $\widetilde g_{h,-i}(\bs_i)$ denotes the provisional kernel estimate at $\bs_i$ obtained after removing the $i$th observation. The selected bandwidth maximises the corresponding leave-one-out Bernoulli log-likelihood,
\begin{equation}\label{eq:M1_cv}
\ell_{\mathrm{CV}}(h)
=
\sum_{i=1}^{n}
\left[
y_i\log\{\hat p_{-i}(h)\}
+
(1-y_i)\log\{1-\hat p_{-i}(h)\}
\right].
\end{equation}

The leave-one-out criterion is evaluated using the provisional kernel estimate \eqref{eq:M1_tildeg}. The Fisher-weighted projection \eqref{eq:M1_orthog} is applied subsequently within each IRLS iteration to define the decomposition between the fixed-effects and smooth components, but it does not alter the spatial neighbourhood represented by the kernel bandwidth.

Because the working response and Fisher weights evolve during the IRLS iterations, the optimal bandwidth also changes. Accordingly, after each IRLS update the leave-one-out criterion \eqref{eq:M1_cv} is re-evaluated and the bandwidth is updated if necessary. This process continues until the selected bandwidth stabilises, after which it is held fixed while the remaining IRLS iterations proceed to convergence. In all analyses considered here, bandwidth stabilisation was rapid, typically occurring within five or six bandwidth updates.

\subsubsection{Mapped residual surface and inference}\label{sec:infer}

At convergence, the smooth component entering the fitted model is the observation-level vector $\widehat{\bg}_{\perp}$, which satisfies the identifying condition $\bX^\top\widehat{\bW}\widehat{\bg}_{\perp}=\mathbf 0$. Spatial interpretation, however, requires a continuously evaluated surface over the study window. Let $\widehat u_i$ and $\widehat w_i$ denote the final values of the partial working residual and Fisher weight defined in \eqref{eq:u} and \eqref{eq:etapw}, respectively. We first define the uncentred location-only kernel surface
\begin{equation}\label{eq:M1_raw_map}
\widehat r_h^{\,\mathrm{raw}}(\bs)
=
\frac{\sum_{i=1}^n\kappa_h(\|\bs-\bs_i\|)\widehat w_i\widehat u_i}
{\sum_{i=1}^n\kappa_h(\|\bs-\bs_i\|)\widehat w_i}.
\end{equation}
Its Fisher-weighted mean over the observed locations is
\[
\widehat c_h
=
\frac{\sum_{i=1}^n\widehat w_i\widehat r_h^{\,\mathrm{raw}}(\bs_i)}
{\sum_{i=1}^n\widehat w_i},
\]
and the mapped residual log-odds surface is defined as
\begin{equation}\label{eq:M1_mapped_surface}
\widehat r_h(\bs)
=
\widehat r_h^{\,\mathrm{raw}}(\bs)-\widehat c_h.
\end{equation}
Consequently, $\sum_{i=1}^n\widehat w_i\widehat r_h(\bs_i)=0$. Its exponentiated form, $\exp\{\widehat r_h(\bs)\}$, represents the local multiplicative departure on the odds scale from the centred covariate-adjusted baseline, with values greater than one indicating locally elevated residual odds.

The observation-level vector $\widehat{\bg}_{\perp}$ is the smooth contribution entering the fitted linear predictor. For mapping, $\widehat r_h(\bs)$ provides a continuous kernel reconstruction of the residual spatial variation represented by the final IRLS working residuals. Positive values indicate locations at which nearby schools exhibit greater case activity than is represented by the fitted fixed-effects component, while negative values indicate the converse. Because the full projection in \eqref{eq:M1_orthog} is defined through the observed design matrix, $\widehat r_h(\bs)$ is not an exact off-sample evaluation of $\widehat g_{\perp}(\bs)$; rather, it is the continuous spatial summary produced by applying the fitted kernel update beyond the observed locations.

To assess whether the mapped residual structure exceeds that expected under a covariate-only explanation, we use a probability-weighted fixed-count label-reassignment scheme. We first fit the covariate-only logistic model and obtain fitted probabilities
\[
\widehat p_i^{(0)}
=
\operatorname{logit}^{-1}
\{\bx_i^\top\widehat{\bbeta}^{(0)}\}.
\]
For each Monte Carlo replicate, exactly $n_1=\sum_{i=1}^n y_i$ observations are selected without replacement to receive case labels, using $\widehat p_i^{(0)}$ as unequal-probability sampling weights; all remaining observations receive control labels. The full orthogonalised spatial model is then refitted to the reassigned labels, with the bandwidth selected from the observed data held fixed, and a centred mapped surface is constructed using \eqref{eq:M1_raw_map}--\eqref{eq:M1_mapped_surface}.

Let $\widehat r_h^{(1)}(\bs),\ldots,\widehat r_h^{(B)}(\bs)$ denote the mapped surfaces from the $B$ Monte Carlo replicates. At each grid location, the upper-tail exceedance proportion is
\begin{equation}\label{eq:M1_exceed}
\widehat\pi_+(\bs)
=
\frac{1}{B}
\sum_{b=1}^B
I\{\widehat r_h^{(b)}(\bs)<\widehat r_h(\bs)\}.
\end{equation}
The corresponding lower-tail quantity is
$
\widehat\pi_-(\bs)
=
\frac{1}{B}
\sum_{b=1}^B
I\{\widehat r_h^{(b)}(\bs)>\widehat r_h(\bs)\}.
$
Large values of $\widehat\pi_+(\bs)$ indicate locations at which the observed surface exceeds most of its covariate-only reference surfaces, while large values of $\widehat\pi_-(\bs)$ indicate unusually low residual risk. Contours of these exceedance proportions identify locations at which the observed residual surface is unusually high or low relative to the covariate-only reference surfaces. The comparisons are location-specific rather than calibrated as a single global test over the entire study window.

A further source of variability arises from random sampling of controls from the eligible school population. As described in Section~\ref{sec:condata}, we repeat the full analysis across $K$ independently sampled control sets. Let $\widehat r_{h_k,k}(\bs)$ denote the mapped surface from control-sampling bag $k$, where $h_k$ is the bandwidth selected for that bag. The bagged mapped surface is the pixelwise average
\begin{equation}\label{eq:M1_surface_pool}
\overline{\widehat{r}(\bs)}
=
\frac{1}{K}
\sum_{k=1}^K
\widehat r_{h_k,k}(\bs),
\end{equation}
and for the Monte Carlo assessment the pooled exceedance proportion is 
\begin{equation}\label{eq:M1_ppool}
\widehat\pi_+^{\mathrm{pool}}(\bs)
=
\frac{1+\sum_{k=1}^K \sum_{b=1}^B
I\{\widehat r_{h_k,k}^{(b)}(\bs)<\widehat r_{h_k,k}(\bs)\}}{1+KB}.
\end{equation}

Coefficient uncertainty is also summarised within and between control-sampling bags. For a single fitted model, the observation-level contribution $\widehat{\bg}_{\perp}$ is treated as a fixed offset in the final logistic mean structure. Writing
$
\widehat p_i
=
\operatorname{logit}^{-1}
\{\bx_i^\top\widehat{\bbeta}+\widehat g_{\perp}(\bs_i)\},
$
and
$
\widehat{\bW}
=
\operatorname{diag}\{\widehat p_i(1-\widehat p_i)\},
$
the conditional model-based covariance matrix is given by
$
(\bX^\top\widehat{\bW}\bX)^{-1}.
$
Across the $K$ control-sampling bags, let $\widehat\beta_{jk}$ denote the estimate of coefficient $j$ from bag $k$, and let
\[
U_{jk}
=
\left[
(\bX_k^\top\widehat{\bW}_k\bX_k)^{-1}
\right]_{jj}
\]
denote its conditional model-based variance. The pooled estimate, average within-bag variance and between-bag variance are
\[
\overline\beta_j
=
\frac{1}{K}\sum_{k=1}^K\widehat\beta_{jk},
\qquad
\overline U_j
=
\frac{1}{K}\sum_{k=1}^K U_{jk},
\qquad
D_j
=
\frac{1}{K-1}
\sum_{k=1}^K
(\widehat\beta_{jk}-\overline\beta_j)^2.
\]
Using a Rubin-style within-between decomposition, the total pooled variance is
\begin{equation}\label{eq:M1_beta_var_pool}
T_j
=
\overline U_j
+
\left(1+\frac{1}{K}\right)D_j.
\end{equation}
Pooled standard errors are obtained as $\sqrt{T_j}$, with Wald confidence intervals and $p$-values formed from $\overline\beta_j$ and $T_j$. These summaries account for both conditional model-based uncertainty within each fitted control set and the additional variability induced by repeated control sampling; their empirical performance is examined in Section~\ref{sec:sim}.

\subsection{Spatio-temporal model}\label{sec:spatiotemporal}

The purely spatial model represents residual variation through a single geographical surface aggregated over the full study period. If the residual risk pattern changes through time, however, this surface may blend spatial structures arising during different periods. We therefore extend the model by allowing the residual field to evolve continuously over both geography and time, avoiding the need to divide the study period into prespecified intervals. A simpler grouped-time formulation is considered in the supplementary materials.

Let $t_i\in\mathbb{R}$ denote the continuous observation time for observation $i$. For cases, this is the recorded incident date. Controls are assigned a continuous time drawn uniformly within their matched calendar year, so that each control represents a school at a contemporaneous point in the study period. The spatio-temporal model is
\begin{equation}\label{eq:M3def}
\operatorname{logit}(p_i)
=
\bx_i^\top\bbeta
+
g_{\perp}(\bs_i,t_i),
\end{equation}
where $g_{\perp}(\bs_i,t_i)$ is the observation-level spatio-temporal contribution entering the fitted linear predictor. Because controls are sampled within calendar year, comparisons are conditioned on the contemporaneous school population and broad changes in the annual frequency of incidents are absorbed by the sampling design. The temporal coordinate in $g_{\perp}$ is therefore used primarily to allow the residual spatial pattern to evolve through time, rather than to estimate an unrestricted national incidence trend.

The orthogonalised IRLS construction of Section~\ref{sec:imp} extends by replacing the spatial kernel with a separable spatio-temporal kernel. With $\eta_i$, $p_i$, $w_i$, $z_i$ and $u_i$ defined as in \eqref{eq:etapw}--\eqref{eq:u}, the provisional smooth at a generic space-time location $(\bs,t)$ is
\begin{equation}\label{eq:M3_tildeg}
\widetilde g_{h,q}(\bs,t)
=
\frac{
\sum_{i=1}^n
\kappa_h(\|\bs-\bs_i\|)
\tau_q(t-t_i)
w_i u_i
}{
\sum_{i=1}^n
\kappa_h(\|\bs-\bs_i\|)
\tau_q(t-t_i)
w_i
},
\end{equation}
where $\kappa_h$ is the isotropic spatial kernel with bandwidth $h$, and
$
\tau_q(v)=q^{-1}\tau(v/q)
$
is a one-dimensional temporal kernel with bandwidth $q$. This separable construction preserves the interpretation of $h$ and $q$ as the spatial and temporal scales of local smoothing, respectively, and is closely related to kernel estimators used in standalone spatio-temporal density and relative-risk estimation \citep[e.g.][]{fernando2014generalizing,davies2018tutorial,davies2019evaluation}.

Let
$
\widetilde{\bg}
=
\left[
\widetilde g_{h,q}(\bs_1,t_1),
\ldots,
\widetilde g_{h,q}(\bs_n,t_n)
\right]^\top.
$
The observation-level contribution is obtained using the same Fisher-weighted projection as in the spatial model,
\begin{equation}\label{eq:M3_orthog}
\bg_{\perp}
=
\widetilde{\bg}
-
\bX(\bX^\top\bW\bX)^{-1}
\bX^\top\bW\widetilde{\bg},
\end{equation}
so that $\bX^\top\bW\bg_{\perp}=\mathbf0$. The regression coefficients are then updated using \eqref{eq:M1_beta_update}, and the working quantities, provisional smooth, projection and coefficient update are iterated until convergence.

Joint data-driven selection of $(h,q)$ would require repeated leave-one-out smoothing over a two-dimensional bandwidth grid and is substantially more demanding than the one-dimensional spatial search. In addition, the resulting criterion may be relatively flat in one or both directions, making stable joint optimization less straightforward in finite samples. Thus, while a spatio-temporal version of \eqref{eq:M1_cv} is a natural selector in principle, in practice one may instead work over a restricted grid of plausible bandwidth pairs or fix $(h,q)$ in advance on pragmatic grounds.


For mapping, the construction in Section~\ref{sec:infer} is extended by replacing the spatial kernel weight $\kappa_h(\|\bs-\bs_i\|)$ with the separable space-time weight $\kappa_h(\|\bs-\bs_i\|)\tau_q(t-t_i)$. Using the final partial working residuals and Fisher weights, this gives the uncentred surface $\widehat r_{h,q}^{\,\mathrm{raw}}(\bs,t)$ and the fit-specific centring constant
\[
\widehat c_{h,q}
=
\frac{
\sum_{i=1}^n
\widehat w_i
\widehat r_{h,q}^{\,\mathrm{raw}}(\bs_i,t_i)
}{
\sum_{i=1}^n\widehat w_i
},
\qquad
\widehat r_{h,q}(\bs,t)
=
\widehat r_{h,q}^{\,\mathrm{raw}}(\bs,t)-\widehat c_{h,q}.
\]
The same centring constant is applied across all time points, rather than recalculating it separately within each time slice, thereby retaining temporal variation in the overall level of the mapped field. The exponentiated surface $\exp\{\widehat r_{h,q}(\bs,t)\}$ gives the corresponding local multiplicative departure on the odds scale.

The mapped surface is evaluated over the study window at selected time slices $t^\star_1,\ldots,t^\star_m$. As in the purely spatial model, $\widehat r_{h,q}(\bs,t)$ is the continuous kernel summary used for mapping, whereas $\widehat{\bg}_{\perp}$ is the Fisher-orthogonal observation-level contribution entering the fitted linear predictor. The same mapping and centring operations are applied to the observed fit and to every Monte Carlo reference fit.

Monte Carlo assessment uses the probability-weighted fixed-count label-reassignment scheme described in Section~\ref{sec:infer}. For each replicate, exactly the observed number of case labels is reassigned without replacement using the covariate-only fitted probabilities as sampling weights. The full spatio-temporal model is then refitted using the fixed bandwidth pair $(h,q)$, and the mapped surface is evaluated at the same time slices as the observed fit. If $\widehat r_{h,q}^{(1)}(\bs,t^\star),\ldots,\widehat r_{h,q}^{(B)}(\bs,t^\star)$ denote the replicate surfaces at time $t^\star$, the upper-tail exceedance proportion is
\begin{equation}\label{eq:M3_exceed}
\widehat\pi_+(\bs,t^\star)
=
\frac{1}{B}
\sum_{b=1}^B
I\left\{
\widehat r_{h,q}^{(b)}(\bs,t^\star)
<
\widehat r_{h,q}(\bs,t^\star)
\right\},
\end{equation}
with the lower-tail quantity defined by reversing the inequality. These comparisons identify locations and times at which the observed mapped surface is unusually high or low relative to the covariate-only reference surfaces. The comparisons are location- and time-specific rather than calibrated as a single global test over the full space-time domain.


The complete spatio-temporal analysis is repeated across the $K$ independently sampled control sets. Mapped surfaces are averaged pixelwise within each time slice, while integer exceedance counts are summed across bags and converted to pooled proportions using the same single plus-one adjustment as in~(\ref{eq:M1_ppool}), with denominator $1+KB$.

\section{Simulation study}\label{sec:sim}

The Fisher-weighted projection in \eqref{eq:M1_orthog} is a defining feature of the proposed fitting scheme, since it explicitly determines how variation shared by the measured predictors and provisional smooth is allocated between the fixed-effects and residual components. Related work has shown that restricted spatial decompositions can have important consequences for coefficient estimation and inference \citep{hodges2010adding,khan2022restricted,khan2026rethinking}. We therefore conduct a focused simulation study comparing the orthogonalised procedure with unrestricted kernel backfitting.
The aim is not to present orthogonalisation as a general remedy for omitted spatial confounding for this class of model, but to assess the practical consequences of the decomposition adopted here.

\subsection{Design}

We generated $n=1500$ locations independently and uniformly over the unit square. Each dataset contained a spatially patterned predictor $x_{\mathrm{sp}}$, formed from a smooth function of location plus independent Gaussian variation, and an independent non-spatial predictor $x_{\mathrm{nsp}}$. Binary responses were generated according to
\begin{equation}\label{eq:sim_dgp}
\operatorname{logit}\{\Pr(Y_i=1)\}
=
\beta_0
+
0.80x_{\mathrm{sp},i}
-
0.45x_{\mathrm{nsp},i}
+
g(\bs_i),
\end{equation}
where $\beta_0$ was calibrated in each replicate to give an expected case prevalence of $1/11$, approximating the case-control ratio in the application. The exact generating functions are provided in the supplementary materials.

We considered four scenarios. In the benchmark scenario (S1), $g(\bs)$ was a smooth location-only field constructed to have little overlap with the spatial trend underlying $x_{\mathrm{sp}}$. The second scenario (S2) deliberately aligned part of $g(\bs)$ with
the broad spatial trend underlying $x_{\mathrm{sp}}$, creating competition between the fixed and smooth components. The final two scenarios introduced an omitted spatial field correlated with the spatial trend in $x_{\mathrm{sp}}$, while varying the amount of non-spatial information in that predictor: one retained substantial independent variation and the other only weak independent variation (S3A and S3B respectively).

Because the coefficient estimand changes under the declared decomposition, we recorded both the coefficients in the original data-generating representation and their Fisher-orthogonal counterparts. Writing $\bg=[g(\bs_1),\ldots,g(\bs_n)]^\top$ and letting $\bW$ contain the true Bernoulli Fisher weights, the latter were defined by
\[
\bbeta_{\perp}
=
\bbeta
+
(\bX^\top\bW\bX)^{-1}\bX^\top\bW\bg,
\qquad
\bg_{\perp}
=
\bg
-
\bX(\bX^\top\bW\bX)^{-1}\bX^\top\bW\bg.
\]
These representations produce the same linear predictor but make explicit the coefficient and residual-field targets associated with orthogonalised fitting. 
Performance was assessed against both the original data-generating
coefficient and the replicate-specific Fisher-orthogonal coefficient
in all four scenarios.

Each dataset was analysed using the proposed orthogonalised kernel estimator, unrestricted kernel backfitting as per \cite{keldig:1998}, a covariate-only logistic regression, a conventional spatial GAM with a thin-plate smooth, and Spatial+ \citep{dupont2022spatialplus}, which first residualises spatially patterned covariates against smooth functions of location. Both kernel procedures used the same fixed bandwidth $h=0.12$. We generated $1000$ replicates under each scenario. Evaluation focused on coefficient bias, root mean squared error, estimated standard errors and coverage of nominal 95\% Wald intervals, together with recovery of the linear predictor and the original and Fisher-orthogonal residual fields.

\subsection{Results}

All methods converged in all simulated datasets. For the declared Fisher-orthogonal target, the proposed estimator had small bias and near-nominal coverage in every scenario (Table~\ref{tab:sim_beta}). Unrestricted kernel backfitting produced similar point estimates in the two scenarios without omitted spatial structure, but its model-based intervals undercover; under omitted spatial structure, both bias and undercoverage became more pronounced. The spatial GAM and Spatial+ estimate different allocations of predictor-aligned spatial variation and consequently departed substantially from the Fisher-orthogonal target in those scenarios.

\begin{table}[t]
\centering
\caption{Bias (coverage) of nominal 95\% intervals for the spatially patterned coefficient from $1000$ replicates.}
\label{tab:sim_beta}
\footnotesize
\setlength{\tabcolsep}{5pt}
\begin{tabular}{@{}lcccc@{}}
\hline
Method & S1 & S2 & S3A & S3B \\
\hline
\multicolumn{5}{l}{\textit{Fisher-orthogonal target $\bbeta_\perp$}}\\
Orthogonal kernel   & $0.020\;(0.951)$ & $0.038\;(0.945)$  & $0.035\;(0.966)$  & $0.020\;(0.961)$ \\
Unrestricted kernel & $0.001\;(0.895)$ & $-0.006\;(0.903)$ & $-0.179\;(0.658)$ & $-0.066\;(0.850)$ \\
Spatial GAM         & $-0.027\;(0.967)$& $-0.093\;(0.922)$ & $-0.290\;(0.544)$ & $-0.260\;(0.775)$ \\
Spatial+            & $-0.016\;(0.953)$& $-0.241\;(0.712)$ & $-0.427\;(0.362)$ & $-0.561\;(0.873)$ \\
Non-spatial GLM     & $0.052\;(0.933)$ & $0.057\;(0.931)$  & $0.091\;(0.905)$  & $0.089\;(0.913)$ \\
\multicolumn{5}{l}{\textit{Original data-generating coefficient
$\bbeta$}}\\

Orthogonal kernel
& $0.049\;(0.932)$
& $0.297\;(0.215)$
& $0.479\;(0.010)$
& $0.628\;(0.000)$ \\

Unrestricted kernel
& $0.030\;(0.891)$
& $0.253\;(0.411)$
& $0.264\;(0.416)$
& $0.542\;(0.021)$ \\

Spatial GAM
& $0.002\;(0.973)$
& $0.166\;(0.827)$
& $0.153\;(0.855)$
& $0.347\;(0.681)$ \\

Spatial+
& $0.013\;(0.954)$
& $0.018\;(0.957)$
& $0.017\;(0.951)$
& $0.047\;(0.957)$ \\

Non-spatial GLM
& $0.081\;(0.887)$
& $0.316\;(0.160)$
& $0.535\;(0.001)$
& $0.697\;(0.000)$ \\
\hline
\end{tabular}
\end{table}




For the original data-generating coefficient, Spatial+ performed well in S2 and S3A, whereas the orthogonalised estimator was more accurate for the Fisher-orthogonal coefficient. This reversal is expected because the methods target different allocations of predictor-aligned spatial
variation; its purpose here is to make that distinction explicit rather than to rank the methods against a single universal estimand.


More importantly, the proposed estimator showed small finite-sample bias and near-nominal coverage for its declared target across all four scenarios, while unrestricted backfitting produced systematically
under-calibrated coefficient intervals.


Recovery of the Fisher-orthogonal residual field was also stable. Across scenarios, the proposed estimator and unrestricted backfitting achieved comparable smooth-field recovery (see Supplementary Table~S3). Thus, orthogonalisation does not materially degrade estimation of the residual field, while its primary effect is to provide a well-defined allocation of shared variation between the fixed and smooth components. Overall linear-predictor recovery was similar across the principal spatial methods, highlighting that their main differences concern how fitted spatial variation is allocated between the fixed and smooth components.

This distinction is important for the present application. Our objective is to estimate covariate-adjusted associations under an explicit allocation of shared spatial variation, while retaining a residual geographical surface for interpretation and mapping. Spatial+ instead estimates the coefficient after removing the smoothly
spatial component of the predictor, so its identifying information comes primarily from the remaining non-spatial variation. When little such variation remains, the resulting estimate may be highly imprecise. The Fisher-orthogonal formulation is therefore more closely aligned with the descriptive inferential objective adopted here.

\section{Application results}\label{sec:res}

We applied the spatial and spatio-temporal models of Section \ref{sec:model} to the school gun-violence data described in Section~\ref{sec:data}, using Gaussian spatial and temporal kernels. Exceedance probabilities were based on $B=199$ Monte Carlo reference fits, and all results were pooled across $K=100$ independently generated control samples. All reported fits satisfied the convergence criteria.


\subsection{Adjusted school-level associations}

The coefficient estimates were very similar under the spatial and spatio-temporal specifications (Table~\ref{tab:betas}). In the spatial model, a one-standard-deviation increase in log enrolment was associated with a $2.36$-fold increase in the adjusted odds of case status (95\% CI: $2.09$-$2.67$). Relative to primary or elementary schools, the corresponding odds were $1.75$ times higher for middle schools (95\% CI: $1.32$-$2.32$) and $7.70$ times higher for high schools (95\% CI: $6.21$-$9.55$). There was little evidence of an association with proportion-male or with the heterogeneous ``other'' school-level category. The linear and quadratic terms for proportion-white and ADI indicated nonlinear adjusted associations. Their fitted shapes are presented using marginal-effect plots in the supplementary materials rather than inferred directly from the individual coefficient signs.

\begin{table*}[t]
\centering
\caption{Pooled coefficient estimates, standard errors, and two-sided p-values under the spatial and spatio-temporal models. Estimates are combined across $K=100$ control samples; standard errors incorporate within- and between-sample variation.}
\label{tab:betas}
\footnotesize
\setlength{\tabcolsep}{7pt}
\begin{tabular}{@{}lrrrr@{}}
\hline
& \multicolumn{2}{c}{Spatial} & \multicolumn{2}{c}{Spatio-temporal} \\
\cline{2-3}\cline{4-5}
Parameter & Est.\ (S.E.) & p-value & Est.\ (S.E.) & p-value \\
\hline
Intercept
& $-4.080\,(0.118)$ & $<0.001$
& $-4.053\,(0.117)$ & $<0.001$ \\

Log enrolment
& $0.861\,(0.062)$ & $<0.001$
& $0.861\,(0.062)$ & $<0.001$ \\

Proportion male
& $0.036\,(0.066)$ & $0.590$
& $0.044\,(0.067)$ & $0.509$ \\

Middle school
& $0.559\,(0.145)$ & $<0.001$
& $0.560\,(0.145)$ & $<0.001$ \\

High school
& $2.041\,(0.110)$ & $<0.001$
& $2.024\,(0.110)$ & $<0.001$ \\

Other school level
& $-0.041\,(0.378)$ & $0.914$
& $-0.048\,(0.378)$ & $0.899$ \\

Proportion white
& $-0.625\,(0.072)$ & $<0.001$
& $-0.608\,(0.073)$ & $<0.001$ \\

Proportion white$^2$
& $0.177\,(0.070)$ & $0.011$
& $0.186\,(0.070)$ & $0.008$ \\

ADI
& $0.195\,(0.047)$ & $<0.001$
& $0.193\,(0.047)$ & $<0.001$ \\

ADI$^2$
& $-0.073\,(0.029)$ & $0.012$
& $-0.071\,(0.029)$ & $0.015$ \\
\hline
\end{tabular}
\vspace{-3mm}
\end{table*}

These estimates describe adjusted associations under the Fisher-orthogonal decomposition and should not be interpreted causally. This caution is particularly important for school racial composition and deprivation, which are better regarded as markers of broader structural and historical processes affecting schools and neighbourhoods. The intercept is likewise not interpreted substantively because its magnitude depends on the case-control sampling design.

The close agreement between the spatial and spatio-temporal coefficient estimates indicates that allowing the residual geographical pattern to evolve over time did not materially alter the adjusted school-level associations. Principal differences between the two specifications will therefore be manifest in the residual risk surfaces.

\subsection{Mapped residual risk}\label{sec:nonparres}

We now examine the fitted surfaces. For the spatio-temporal model, an animation and a rotatable three-dimensional iso-surface visualisation are also provided in the supplementary materials.

\subsubsection{Spatial pattern}

Bandwidth selection was performed separately for each of the $K=100$ control samples. The mean selected bandwidth was $160.32$ km, with a mean $\pm2$ standard-deviation range of $115.1$-$205.6$ km.

\begin{figure}[h!]
    \centering
    \includegraphics[width=0.8\textwidth,trim=0 0 40 20,clip]{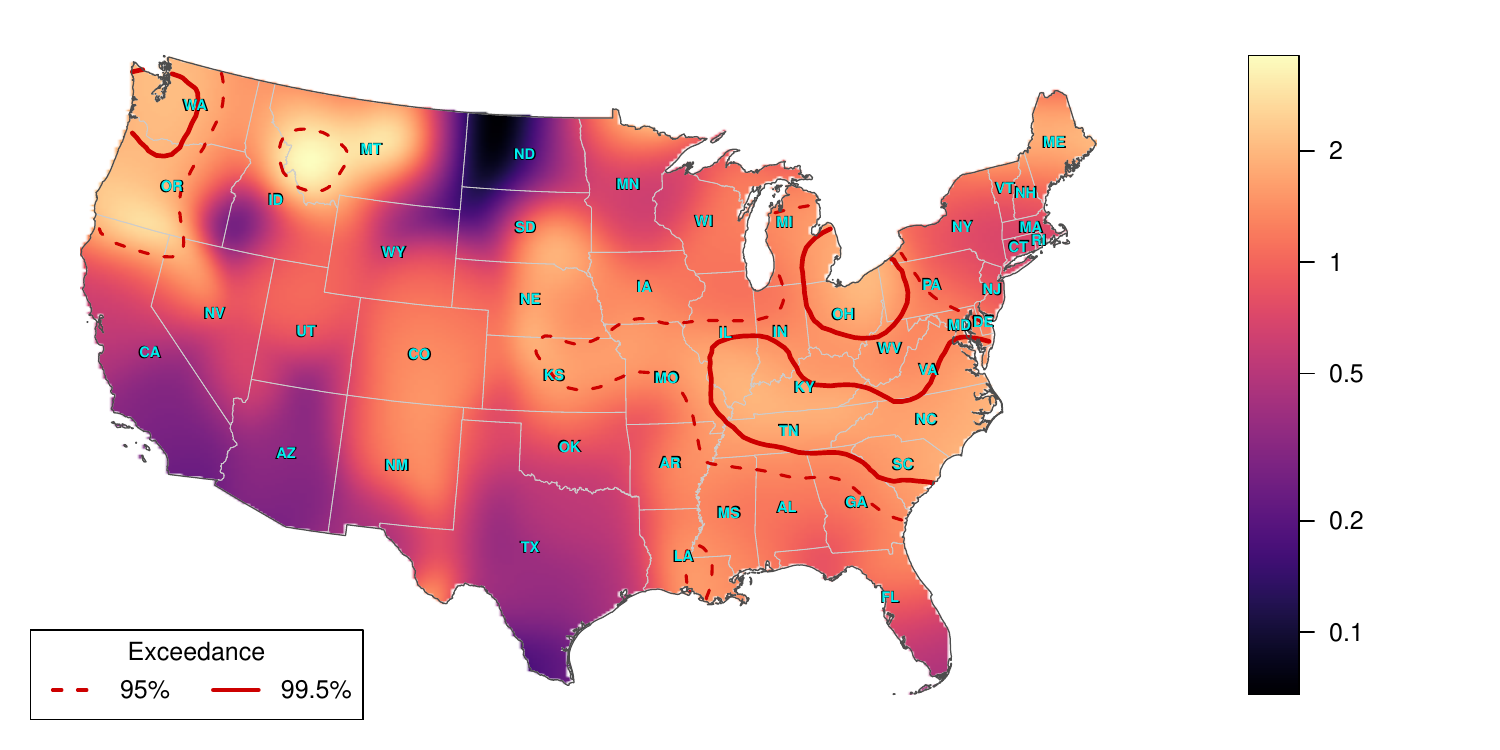}
    \vspace{-4mm}
    \caption{Bag-averaged mapped residual odds multiplier for the purely spatial model. Dashed and solid lines denote pooled location-specific exceedance probabilities of $0.95$ and $0.995$, respectively; logarithmic colour scale.}
    \label{fig:M1g}
    \vspace{-1mm}
\end{figure}

Figure~\ref{fig:M1g} shows $\exp\{\overline{\widehat r(\bs)}\}$, obtained by exponentiating the bag-averaged mapped log-odds residual surface of (\ref{eq:M1_surface_pool}). Values above one indicate locations where nearby schools exhibit more case activity than is represented by the fitted school-level covariates, whereas values below one indicate less. The superimposed contours mark aggregated location-specific exceedance probabilities of $0.95$ and $0.995$ arising from (\ref{eq:M1_ppool}).

The dominant feature is a broad corridor of elevated residual risk through the central and eastern United States, with the strongest contiguous exceedance region extending through the lower Midwest, Kentucky and Tennessee, and into Appalachia and the southeastern interior. Thus, even after adjustment for measured school characteristics, case occurrence remains locally elevated across this corridor.

By contrast, much of California, the Southwest and southern Texas has a mapped multiplier below one and little corresponding exceedance evidence. Because cases are compared with controls drawn from the background school population, it represents a local departure from baseline after accounting for both measured school characteristics and the geographical distribution of schools. Smaller elevated regions are also visible in the Pacific Northwest, northern Rocky Mountains and around the Great Lakes. Overall, the unexplained component is geographically structured rather than diffuse.

As a targeted diagnostic, we compared the orthogonalised fit with unrestricted kernel backfitting using the same bandwidth within each control sample. The broad geographical pattern was similar, but the unrestricted smooth showed substantial overlap with the fixed-effects space and produced systematic coefficient shifts. Orthogonalisation therefore primarily affected the allocation of shared spatial variation between the fixed and smooth components; full results are provided in the supplementary materials.

\subsubsection{Spatio-temporal pattern}

The spatio-temporal model was fitted using spatial and temporal bandwidths of $h=230$ km and $q\approx 548$ days (1.5 years), respectively. Exploratory fits using bandwidths scaled by $0.75$ and $1.5$ produced the expected changes in granularity but did not materially alter the dominant residual pattern or the coefficient estimates.

\begin{figure}[h!]
    \vspace{-0mm}
    \centering
    \includegraphics[width=0.3\textwidth]{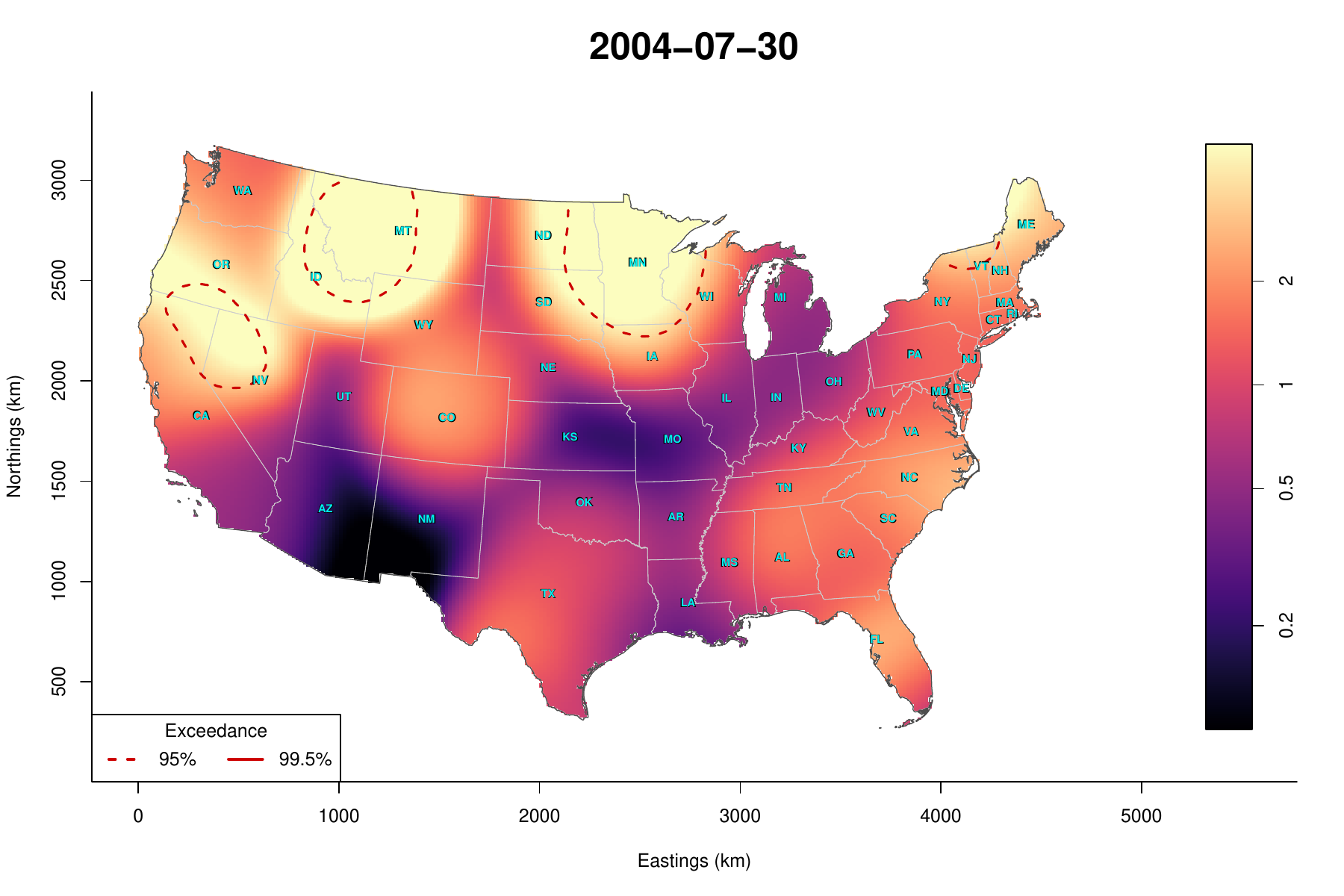}
    \includegraphics[width=0.3\textwidth]{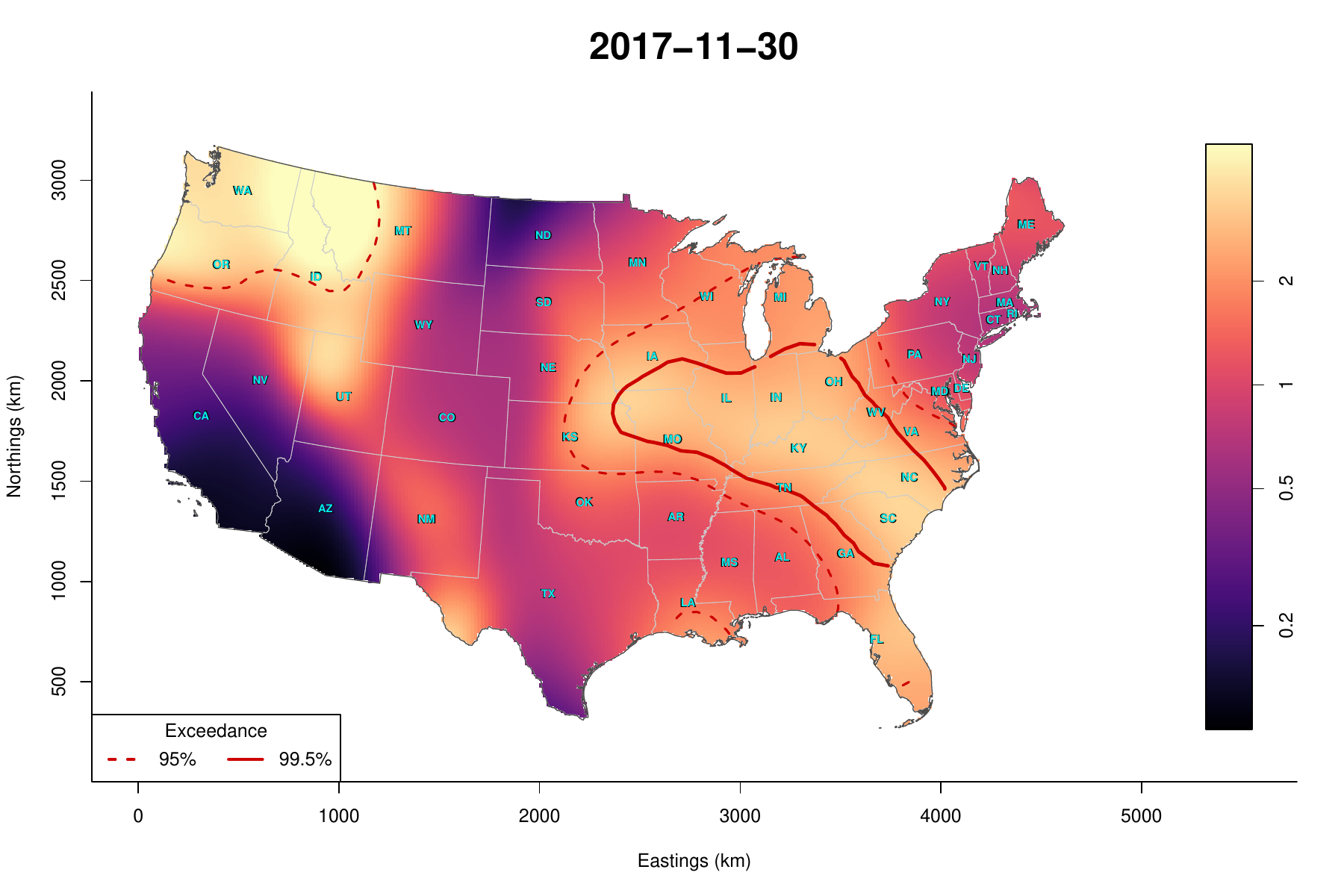}
    \includegraphics[width=0.3\textwidth]{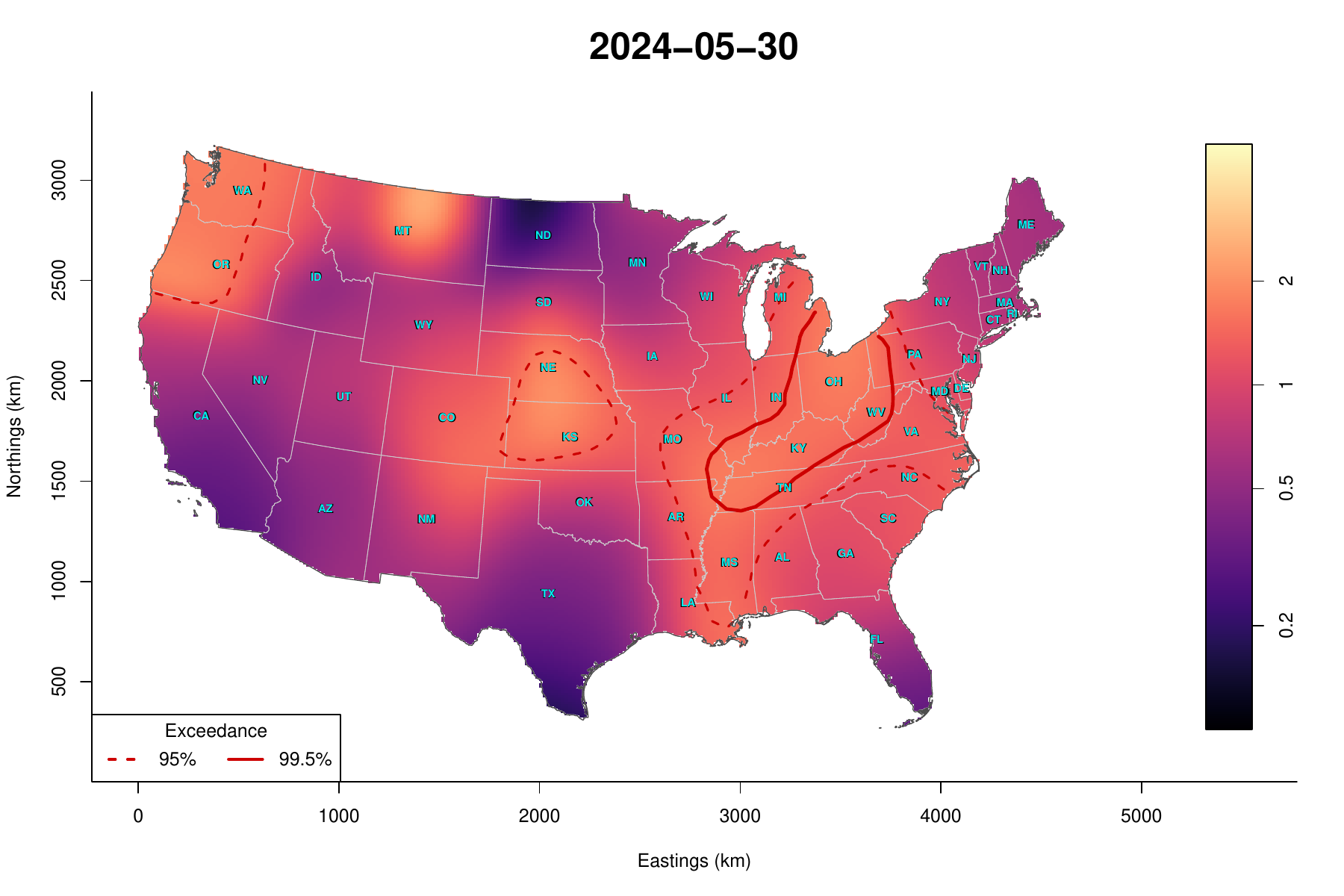}\\
    \includegraphics[width=0.48\textwidth,trim=0 -60 0 0,clip]{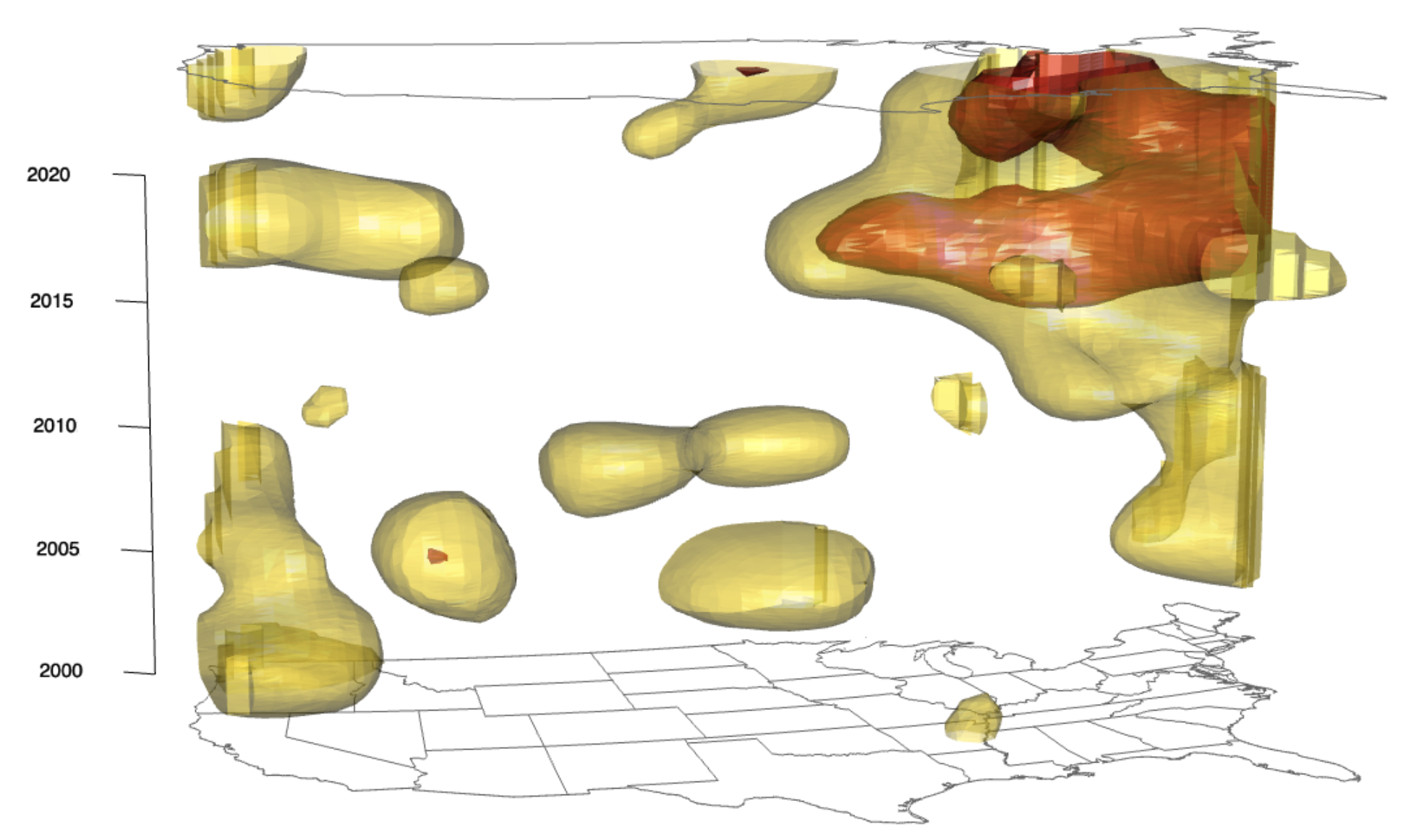}
    \includegraphics[width=0.42\textwidth]{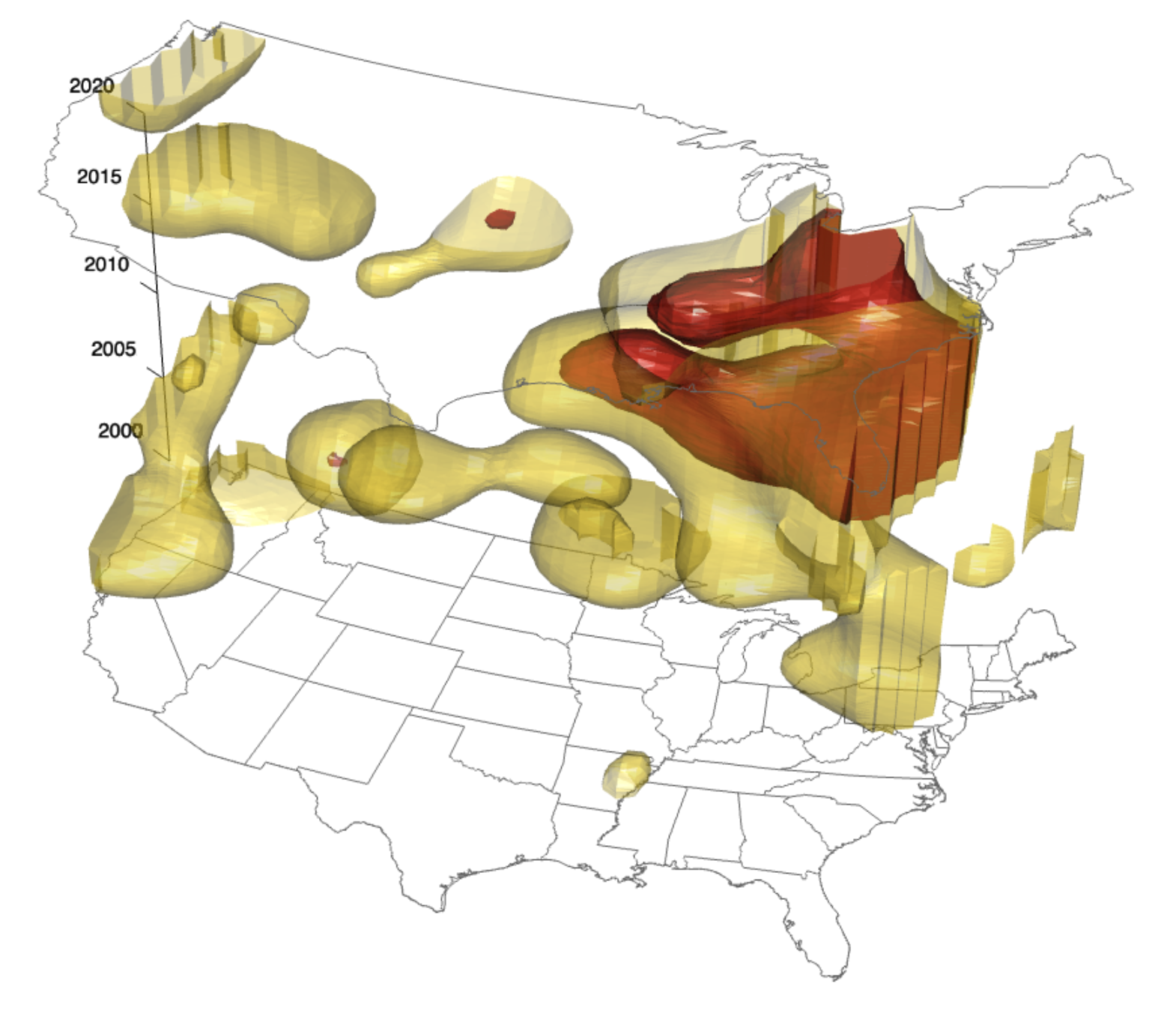}
    \vspace{-3mm}
    \caption{Selected frames of the bag-averaged spatio-temporal mapped residual odds multiplier (top), with dashed and solid contours denoting pooled location-specific exceedance probabilities of $0.95$ and $0.995$, respectively. The lower panels show two views of the corresponding space-time exceedance iso-surfaces.}
    \label{fig:M3g}
    \vspace{-1mm}
\end{figure}

The bag-averaged mapped residual odds multiplier,
$\exp\{\overline{\widehat r_{h,q}(\bs,t)}\}$, is most naturally viewed as an animation, provided in the supplementary materials and online at
\footnotesize\url{https://www.stats.otago.ac.nz/research/davies-gv/animation.mp4}\normalsize.
Because the same fit-specific centring constant is used at every time point, temporal changes in the overall mapped level are retained. Figure~\ref{fig:M3g} shows three representative frames together with two views of the trivariate $0.95$ and $0.995$ exceedance iso-surfaces as per (\ref{eq:M3_exceed}); an interactive version is available at
\footnotesize\url{https://www.stats.otago.ac.nz/research/davies-gv/exceed.html}\normalsize.

The dominant later-period feature is a connected central-eastern region of elevated residual risk extending through parts of the Midwest, Appalachia and the Southeast. The broader $0.95$ exceedance region becomes clearly established by the mid-2010s and persists through much of the remainder of the study period, while the more concentrated $0.995$ core is most prominent in the later years. The space-time connectivity of these regions suggests an evolving but sustained central-eastern residual-risk pattern, rather than a succession of wholly unrelated local peaks.

Earlier and western features are generally smaller or more detached in space-time, consistent with more localised or episodic residual elevation. These include recurring features in the Pacific Northwest and a pronounced central-interior pattern around the late 2000s and early 2010s. The spatio-temporal fit therefore refines the spatial result: the dominant central-eastern pattern emerges most clearly in the latter portion of the study period and subsequently changes in extent and intensity, alongside regional features with more limited durations.

\section{Concluding remarks}\label{sec:disc}

This study refines a semiparametric case-control framework for jointly estimating adjusted school-level associations and mapping residual variation across space and time. A notable feature of the implementation is its Fisher-weighted orthogonalisation. This makes the allocation of shared variation explicit while retaining a continuous residual surface for interpretation. 

Applied to school gun violence incidents across the contiguous United States, the analysis identified strong and stable associations with school size and level. Larger schools, and especially high schools, had substantially greater adjusted odds of case status than otherwise comparable primary or elementary schools. The spatial and spatio-temporal fits gave closely similar coefficient estimates, indicating that these associations were not materially altered by allowing the residual geographical pattern to evolve over time.

The mapped residual component nevertheless revealed substantial structure beyond the measured school characteristics. The dominant feature was a broad central-eastern corridor of elevated residual risk, extending through parts of the Midwest, Appalachia and the Southeast. The spatio-temporal fit suggested that this pattern became clearly established by the mid-2010s and persisted through much of the remainder of the study period, becoming most pronounced in later years, while smaller western and interior features were generally more localised or episodic. These findings indicate that the national pattern is associated not only with the composition of individual schools, but also with broader geographically organised conditions not represented by the available predictors.

Several qualifications are important. The coefficients describe adjusted associations rather than causal effects, and the mapped residual surface is a model-derived, smoothed summary rather than a direct measurement of an underlying risk process. The analysis is also limited by the coverage and classification of the incident database, the restriction to public schools, and the school-level variables available nationally. The exceedance contours provide location-specific comparisons with covariate-only reference fits and should not be interpreted as a single globally calibrated hypothesis test.

A natural next step is to examine whether the persistent residual pattern can be related to broader contextual variables, such as firearm-policy environments, violent-crime conditions and socioeconomic characteristics. Such analyses would be associational and should explicitly accommodate the fact that the response is itself a spatially smoothed quantity. Nevertheless, they may help identify which regional conditions warrant closer substantive investigation. More generally, the results show the value of treating mapped residual structure as a primary scientific output, rather than merely as a nuisance component of the fitted model.

\section*{Data and code availability (peer review)}
For the purposes of peer review, the data and code supporting the application findings are provided at \fbox{\url{https://doi.org/10.6084/m9.figshare.33136094}}.
See Section S6 of the supplementary PDF document for instructions on use.

\section*{Competing interests}
No competing interest is declared.

\section*{Author contributions statement}
T.M.D. conceived the project, curated and prepared the data, developed the code, conducted the analyses, interpreted the findings, and drafted the manuscript. M.R.D., A.H.\ and G.W. contributed substantive contextual expertise, assisted with interpretation of the findings, and critically reviewed and revised the manuscript.


\section*{Acknowledgments}
This research and its findings would not have been possible without the ongoing data collection and management efforts of Dr.\ David Riedman and the School Shooting Database. T.M.D. acknowledges financial support from the Royal Society Te Ap\={a}rangi (Marsden Fund Grant no.\ 23-UOO-148), and RSL/sabbatical funding from U.\ Otago.

\bibliographystyle{abbrvnat}
\bibliography{literature}

\end{document}